\documentclass[preprint,letterpaper,prc,showpacs]{revtex4}
\usepackage{graphicx}
\newcommand{\beq}{\begin{equation}}
\newcommand{\eeq}{\end{equation}}
\newcommand{\beqa}{\begin{eqnarray}}
\newcommand{\eeqa}{\end{eqnarray}}
\newcommand{\nn}{\nonumber}
\newcommand{\Sigs}{\Sigma_{\mathrm s} }
\newcommand{\Sigv}{\Sigma_{\mathrm v} }
\newcommand{\Sigo}{\Sigma_{\mathrm o} }
\newcommand{\Sigsn}{\Sigma_{{\mathrm s},n} }

\newcommand{\Sigon}{\Sigma_{{\mathrm o},n} }
\newcommand{\Sigsp}{\Sigma_{{\mathrm s},p} }

\newcommand{\Sigop}{\Sigma_{{\mathrm o},p} }
\newcommand{\kf}{k_{\mathrm F} }
\newcommand{\kfn}{k_{\mathrm Fn} }
\newcommand{\kfp}{k_{\mathrm Fp} }
\newcommand{\bfgamma}{\mbox{\boldmath$\gamma$\unboldmath}}
\newcommand{\veck}{\textbf{k}}
\newcommand{\vecp}{\textbf{p}}
\newcommand{\vecq}{\textbf{q}}
\newcommand{\vecu}{\textbf{u}}

\newcommand{\pabs}{|{\bf p}|}
\newcommand{\qabs}{|{\bf q}|}
\begin{document}
\preprint{}
\title{The Relativistic Dirac-Brueckner Approach to Asymmetric Nuclear Matter}
\author{E. N. E. van Dalen}
\author{C. Fuchs}
\author{Amand Faessler}
\affiliation{Institut
f$\ddot{\textrm{u}}$r Theoretische Physik, Universit$\ddot{\textrm{a}}$t
T$\ddot{\textrm{u}}$bingen,
Auf der Morgenstelle 14, D-72076 T$\ddot{\textrm{u}}$bingen, Germany}
\begin{abstract}
The properties of asymmetric nuclear matter have been investigated in
a relativistic Dirac-Brueckner-Hartree-Fock framework using the Bonn A
potential. The components
of the self-energies are extracted by projecting on Lorentz invariant
amplitudes. Furthermore, the optimal representation scheme for the $T$
matrix, the subtracted $T$ matrix representation, is applied and the
results are compared to those of other representation schemes.
Of course, in the limit of symmetric nuclear matter our results agree with those
found in literature. 
The binding energy $E_b$ fulfills the quadratic dependence on the
asymmetry parameter and the symmetry energy is 34 MeV at saturation
density. Furthermore, a neutron-proton effective mass splitting of
$m_n^* < m_p^*$ is found. In addition, results are given for 
the mean-field effective coupling constants.  
\end{abstract}
\maketitle
%
\section{Introduction}
%
Symmetric nuclear matter has been studied extensively. The 
conventional nonrelativistic approach to nuclear matter, the BHF
(Brueckner-Hartree-Fock) theory,  goes back to earlier works
by Brueckner and
others~\cite{brueckner54,bethe56,goldstone57,bethe71,haftel70,sprung72}. 
A breakthrough was achieved when the first relativistic (Dirac-)
Brueckner-Hartree-Fock (DBHF) calculations were performed in the
eighties~\cite{anastasio83,horowitz87,brockmann90}. It could
describe remarkably successfully the saturation
properties of nuclear matter, saturation energy and density
of the equation of state (EOS).
\\ \indent Investigations of asymmetric nuclear matter, however, are
rather rare. The breaking of isospin-symmetry complicates the problem
considerably compared to symmetric or pure neutron matter. 
Some older studies within the
nonrelativistic Brueckner scheme can be
found in Refs.~\cite{brueckner68,siemens70}. In
Ref.~\cite {bombaci91} 
a calculation  in the nonrelativistic Brueckner scheme is
presented for the Paris potential. Furthermore, 
the properties of isospin-asymmetric nuclear matter have been
investigated in the framework of an extended nonrelativistic Brueckner approach in Ref.~\cite{zuo99}.
In addition, relativistic
Brueckner calculations are performed in Ref.~\cite{terhaar87b,engvik94,engvik96,huber95,dejong98,lee98,schiller01,alonso03}. Finally, we
mention two calculations
within the $\sigma-\omega$ model~\cite{prakash87,frohlich98}. 
\\ \indent The investigation of asymmetric
matter is of importance for astrophysical studies such as the physics of supernova
explosions~\cite{bethe90} and of neutron
stars~\cite{pethick95},  neutron-rich
nuclei~\cite{tanihata95,dieperink03}, and their collisions~\cite{li97}. 
The interest for the structure of  neutron-rich
nuclei and the collisions between these nuclei is only of recent date, because data for
asymmetric nuclei were scarce in the past. However, this situation
changes with recent advances in the development of high-intensity radioactive beam
facilities that will produce nuclei with large neutron excess. 
Hence,
systematic theoretical studies of asymmetric nuclear matter are
becoming more important. 
\\ \indent An important issue is the
determination of the precise form of the nucleon self-energy. To
determine the Lorentz
structure and momentum dependence of the self-energy, the positive-energy-projected in-medium
on-shell $T$ matrix has to be decomposed into Lorentz invariant
amplitudes. Because of  the restriction to positive energy states 
ambiguities~\cite{nuppenau89} arise, because pseudoscalar ($ps$) and
pseudovector ($pv$) components
can not uniquely be disentangled for on-shell scattering. However,
with a pseudoscalar vertex the pion couples maximally to negative
energy states which are not included in the standard Brueckner
approach. This coupling to negative energy states is inconsistent
with the potentials used and leads to large and
spurious nuclear self-energy contributions from negative energy states~\cite{sehn97}. Hence, it
was further demonstrated~\cite{fuchs98} that the conventional $pv$ representation used
to cure this problem fails. The reason is that pseudoscalar admixtures
are still present. Finally, new and reliable methods, the complete $pv$
representation~\cite{fuchs98} and the subtracted $T$ matrix
representation~\cite{gross99}, were
proposed to remove those spurious contributions from the
$T$ matrix  for symmetric nuclear matter.
In contrast, only the conventional $pv$ representation
has been applied for asymmetric nuclear matter~\cite{dejong98}.
\\ \indent In this work we describe asymmetric nuclear  matter at zero
temperature in the relativistic DBHF approach using the Bonn
potential and their bare $NN$ matrix elements $V$~\cite{machleidt89}. Furthermore, the optimal representation scheme for the $T$ matrix, the subtracted $T$
matrix representation, is applied. A comparison with other
representation schemes is made, such as the $ps$ representation, the
conventional $pv$ representation, and the
complete $pv$ representation. In addition, the relativistic Pauli
operator is used. Compared to symmetric nuclear
matter the theoretical and numerical effects are larger because
protons and neutrons are occupying different Fermi
spheres. 
\\ \indent The plan of this paper is as follows. The relativistic DBHF
is discussed in Sec.~\ref{sec:RBA}. Furthermore, Sec.~\ref{sec:CR&SE} is devoted
to the covariant representation of the in-medium $T$ matrix in connection with nucleon self-energy
components. The results are presented and discussed in
Sec.~\ref{sec:R&D}. Finally, we end with a summary and a conclusion in
Sec.~\ref{sec:c}.
%
\section{Relativistic Brueckner Approach}
\label{sec:RBA}
In this section the relativistic Brueckner approach is discussed.
First a short description of the relativistic Brueckner approach  is
given for symmetric nuclear matter. Next the modifications for the
asymmetric case will be treated. 
%
\subsection{Symmetric Nuclear Matter}
%
In the relativistic Brueckner approach the nucleon inside 
nuclear matter may be viewed as a dressed particle 
as a consequence of its two-body interaction with the surrounding
nucleons. The in-medium interaction of the nucleons is treated in the ladder
approximation of the relativistic Bethe-Salpeter (BS) equation
\beqa
T = V + i \int  V Q G G T,
\label{subsec:SM;eq:BS}
\eeqa
where $T$ denotes the $T$ matrix and $V$ is the bare nucleon-nucleon
interaction. The intermediate off-shell nucleons are described by a two-body propagator $iG G$.
The Pauli operator Q accounts for the 
influence of the medium by the Pauli principle and prevents scattering
to occupied states.
The Green's function $G$ which describes the propagation of dressed
nucleons in the medium fulfills the Dyson equation
\beq
G=G_0+G_0\Sigma G.
\label{subsec:SM;eq:Dysoneq}
\eeq
$G_{0}$ denotes the free nucleon propagator while the influence of the 
surrounding nucleons is expressed by the self-energy $\Sigma$. 
In Brueckner formalism  this self-energy is determined by summing up the
interactions with all the nucleons inside the Fermi sea $F$ in
Hartree-Fock approximation   
\beqa
\Sigma = -i \int\limits_{F} (Tr[G T] - GT ).
\label{subsec:SM;eq:HFselfeq1}
\eeqa
The coupled set of 
Eqs.~(\ref{subsec:SM;eq:BS})-(\ref{subsec:SM;eq:HFselfeq1})
represents a self-consistency problem and has to be iterated until
convergence is reached. Below the approximations which are made in
the standard relativistic Brueckner approach to solve the coupled set
of Eqs.~(\ref{subsec:SM;eq:BS})-(\ref{subsec:SM;eq:HFselfeq1}) 
are discussed.
%
\subsubsection{Self-consistent spinor basis}
%
Because of the requirement of translational and rotational invariance,
hermiticity, parity conservation, and time reversal invariance,
the most general form of the Lorentz structure of the self-energy  in
the nuclear matter rest frame is   
\beqa
\Sigma(k,\kf)= \Sigs (k,\kf) -\gamma_0 \, \Sigo (k,\kf) + 
\bfgamma  \cdot \textbf{k} \,\Sigv (k,\kf),
\label{subsec:SM;eq:self1}
\eeqa
The $\Sigs$, $\Sigo$, and $\Sigv$ components are Lorentz scalar
functions which actually depend on the Lorentz invariants $k^2$,$k
\cdot  j$ and $j^2$, where $j_{\mu}$ denotes the baryon
current. Hence, the Lorentz invariants can be expressed in terms of
$k_0$, $|\veck|$ and $\kf$, where $\kf$ denotes the Fermi momentum.
The components of the self-energy are easily determined by taking the
respective traces \cite{horowitz87,sehn97}
\beq
\Sigs = \frac{1}{4} tr \left[ \Sigma \right],\quad 
\Sigo = \frac{-1}{4} tr \left[ \gamma_0 \, \Sigma \right], \quad 
\Sigv =  \frac{-1}{4|\veck|^2 } 
tr \left[{\bfgamma}\cdot \veck \, \Sigma \right]. 
\label{subsec:SM;eq:trace}
\eeq
\\ \indent The presence of the medium leads to effective masses and effective
momenta of the nucleons
\beqa
m^*(k, \kf) = M + \Re e \Sigma_s(k, \kf), \quad k^{*\mu}=k^{\mu} + \Re e
\Sigma^{\mu}(k, \kf).
\label{subsec:SM;eq:dirac}
\eeqa
The Dirac equation written in terms of these effective masses and
momenta has the form
\beq
 [ \gamma_\mu k^{*^\mu} - m^*(k,\kf)] u(k,\kf)=0.
\label{subsec:SM;eq:dirac1}
\eeq
In the following we will work in the quasi-particle approximation,
i.e. the imaginary part of the self-energy $\Im m \Sigma$ will be
neglected, to simplify  the self-consistency scheme.
To determine the self-energy only positive-energy states are
needed in the relativistic Brueckner approach. Therefore,
the full nucleon propagator can be replaced in
Eq.~(\ref{subsec:SM;eq:HFselfeq1}) by
the Dirac propagator \cite{horowitz87,serot86}
\beqa
G_D(k, \kf)=[\gamma_{\mu} k^{* \mu} + m^*(k, \kf)] 2 \pi i \delta(k^{* 2} - m^{*
  2}(k, \kf)) \Theta(k^{*0}) \Theta(k_F-|\veck|).
\eeqa
Here $k$ denotes the momentum of a nucleon inside 
the Fermi sea in the nuclear matter rest frame. 
This momentum is on mass shell. 
Due to the $\Theta$-functions in the propagator only positive energy nucleons
are allowed in the intermediate scattering states which eliminates the
divergent contributions from negative energy states. 
Thus we avoid the delicate problem of infinities in the
theory which generally will occur if one includes contributions 
from negative energy nucleons in the Dirac sea 
\cite{horowitz87,dejong98,poschenrieder88}. 
\\ \indent By the introduction of the reduced effective mass and kinetic
momentum
\beq
{{\tilde k}^*}_\mu = k^*_\mu / \left( 1+\Sigv(k,\kf)\right) \, , \,
{\tilde m}^*(k,\kf) = m^*(k,\kf)/ \left( 1+\Sigv(k,\kf)\right),
\label{subsec:SM;eq:redquantity}
\eeq
the Dirac equation in the nuclear matter rest frame can be rewritten
as
\beq
 [ \gamma_\mu {\tilde k}^{*^\mu} - {\tilde m}^*(k,\kf)] u(k,\kf)=0.
\quad 
\label{subsec:SM;eq:dirac2}
\eeq
In general the reduced effective mass is density and momentum
dependent. To simplify the calculation, however, 
the ``reference spectrum approximation''~\cite{bethe63} is applied in the
iteration procedure, i.e. the effective
mass of the nucleon is assumed to be entirely density dependent
($|\veck|=k_F$). The method implies that the self-energy 
itself is only weakly momentum dependent. 
Therefore, at the end of the calculation one has to verify the consistency of the 
assumption $\Sigma(k)\approx \Sigma(|\veck|=\kf)$ 
with the result of the iteration procedure.
\\ \indent The solution of the Dirac equation in Eq.~(\ref{subsec:SM;eq:dirac2}) provides the in-medium nucleon
spinor 
\beq
u_\lambda (k,\kf)= \sqrt{ { {\tilde E}^*(\veck)+ {\tilde m}^*_F}\over
{2{\tilde m}^*_F}} 
\left( 
\begin{array}{c} 1 \\ 
{2\lambda |\veck|}\over{{\tilde E}^*(\veck)+ {\tilde m}^*_F}
\end{array}
\right)
\chi_\lambda,
\eeq
where ${\tilde E}^*(\veck)=\sqrt{\veck^2+{\tilde m}^{*2}_F}$
\footnote{From now on we omit the tilde because in the following we
  normally deal with ${\tilde m}^*_F,{\tilde k}^{*^\mu}$.}.
$\chi_\lambda$ denotes a two-component Pauli spinor with 
$\lambda=\pm {1\over 2}$. The normalization of the Dirac spinor is 
thereby chosen as $\bar{u}_\lambda(k,\kf) u_\lambda(k,\kf)=1$. 
Since the spinor contains the reduced effective 
mass the matrix elements of the 
bare nucleon-nucleon interaction become density dependent. 
This density dependence, which is absent in nonrelativistic Brueckner
calculations, is actually considered as the main reason for 
the success of the relativistic DBHF approach concerning 
the description of the nuclear saturation mechanism
\cite{anastasio83}.  
%
\subsubsection{Covariant $T$ matrix}
%
We apply the relativistic Thompson equation \cite{thompson70} to solve the 
scattering problem of two nucleons in the nuclear medium. Therefore the
two-particle propagator $iGG$ in the BS equation, Eq.~(\ref{subsec:SM;eq:BS}), has to be replaced by
the effective Thompson propagator~\cite{thompson70}. 
The Thompson propagator implies that the time-like component of the
momentum transfer in $V$ and $T$ is set equal to zero which is a
natural constraint in the c.m. frame, but not a covariant one.
The Thompson propagator projects the intermediate nucleons onto
positive energy states and restricts the exchanged energy transfer by
$\delta(k^0)$ to zero. Thus Eq.~(\ref{subsec:SM;eq:BS}) is reduced to a three-dimensional
integral equation. In the two-particle center of mass (c.m.) 
frame - the natural frame for studying two-particle 
scattering processes - the Thompson equation can be written as~\cite{terhaar87a,sehn97}
\beqa
T(\vecp,\vecq,x)|_{c.m.} &=&  V(\vecp,\vecq) 
\label{subsec:SM;eq:thompson} \\
&+& 
\int {d^3\veck\over {(2\pi)^3}}
{\rm V}(\vecp,\veck)
{m^{*2}_F\over{E^{*2}(\veck)}}
{{Q(\veck,x)}\over{2{E}^*(\vecq)-2{E}^*(\veck)
+i\epsilon}}
T(\veck,\vecq,x), 
\nonumber
\eeqa
where $\vecq=(\vecq_1 - \vecq_2)/2$ is the relative three-momentum 
of the initial state and $\veck$ and $\vecp$ are the relative 
three-momenta of the intermediate and the final states, respectively.
The Pauli operator $Q$ depends on the boost three velocity $\vecu$ into
the c.m. frame. Hence, the $T$ matrix depends on the set of parameters $x=\{\kf,m^*_F,|\vecu|\}$.
\\ \indent  The Thompson equation~(\ref{subsec:SM;eq:thompson}) for
the on-shell $T$ matrix ($|\vecp|=|\vecq|$) can be solved applying
standard techniques described in detail by Erkelenz~\cite{erkelenz74}. Doing
so, the positive-energy helicity $T$ matrix elements are determined explicitly
via the $|JMLS>$-scheme. In the on-shell case only five, for asymmetric
nuclear matter six, of the sixteen helicity
matrix elements are independent which follows from general
symmetries~\cite{erkelenz74}.
After a partial wave projection onto the $|JMLS>$-states the integral
reduces to an one-dimensional integral over the relative momentum
$|\veck|$ and Eq.~(\ref{subsec:SM;eq:thompson}) decouples into three
subsystems of integral
equations: the uncoupled spin singlet, the uncoupled spin triplet,
and the coupled triplet states. To achieve the reduction to the
one-dimensional integral equations the Pauli operator
$Q$ has to be replaced by an angle-averaged Pauli operator
$\overline{Q}$ ~\cite{horowitz87}. Since the Fermi sphere is deformed to a
Fermi ellipsoid in the two-nucleon c.m. frame, $\overline{Q}$ is
evaluated for such a Fermi ellipsoid:
\beqa
\overline{Q} = \left\{ \begin{array}{c c c}
0                                                             &     & |\veck|<k_{min} \\
\frac{\gamma E^*(k)-E^*_{F}}{\gamma u | \veck | }     & for & k_{min}< | \veck | <k_{max} \\
1                                                             &     & k_{max} < |\veck| 
\end{array} \right.
\label{subsec:SNM;eq:Pauli}
\eeqa
with $k_{min}=\sqrt{\kf^2- u^2 E_F^2}$, 
$k_{max}=\gamma (u E_{F}+ \kf)$, and $u=|\vecu|$. The integral equations are solved
by the matrix inversions
techniques of Haftel and Tabakin~\cite{haftel70}.
\\ \indent The two-nucleon states are two-fermion states. Due to the
anti-symmetry of these states the total isospin of the two-nucleon
system $({\rm I}=0,1)$ can be restored by the standard
selection rule 
\beq
(-1)^{\rm L+S+I}=-1.
\label{selection}
\eeq 
With help of Eqs.~(3.28) and~(3.32) in~\cite{erkelenz74}  the five independent partial wave
amplitudes in the helicity representation are obtained from the five independent on-shell amplitudes in the
$|JMLS>$-representation. The summation over the total angular momentum yields 
the five on-shell plane-wave helicity matrix elements 
\beqa
<{\vecp} \lambda_1^{'} \lambda_2^{'}| T^{\rm I}(x)|
{\vecq} \lambda_1 \lambda_2>
= \sum\limits_{\rm J} \left( \frac{2{\rm J}+1}{4\pi}\right) 
d^J_{\lambda^{'} \lambda}(\theta)
<\lambda_1^{'} \lambda_2^{'}| T^{\rm J,I}(\vecp,\vecq,x)|
\lambda_1 \lambda_2> .
\nonumber \\
\label{tmatel1}
\eeqa
Here $\theta$ is the scattering angle between $\bf q$ and $\bf p$, with
$\pabs=\qabs$, while $\lambda=\lambda_1-\lambda_2$ and 
$\lambda^{'}=\lambda_1^{'}-\lambda_2^{'}$.
The reduced rotation matrices $d^{\rm J}_{\lambda^{'} \lambda}(\theta)$ 
are those defined by  Rose \cite{rose57}.
%
\subsection{Asymmetric Nuclear Matter}
%
In symmetric nuclear matter the Fermi momenta of protons and
neutrons are equal. However, in asymmetric nuclear matter the protons
and neutrons are occupying different Fermi spheres, leading to
different Pauli operators and corresponding neutron and
proton effective masses and self-energies. This asymmetry is also
reflected in the fact that one has three different in-medium interactions
of the nucleons. They are treated in the ladder
approximation of the relativistic Bethe-Salpeter equation
\beqa
T_{nn} = V_{nn} + i \int  V_{nn} Q_{nn} G_{n} G_{n} T_{nn}, \label{sec:RBAAM;eq:BSeqnn}  \\
T_{pp} = V_{pp} + i \int  V_{pp} Q_{pp} G_{p} G_{p} T_{pp}, \label{sec:RBAAM;eq:BSeqpp}  \\
T_{np}^{dir} = V_{np}^{dir} + i \int  V_{np}^{dir} Q_{np}
G_{n} G_{p} T_{np}^{dir} + i \int  V_{np}^{exc}
Q_{pn} G_{p} G_{n} T_{np}^{exc},
\label{sec:RBAAM;eq:BSeqnpdir}  
\eeqa
and
\beqa 
T_{np}^{exc} = V_{np}^{exc} + i \int  V_{np}^{exc}  Q_{pn}
G_{p} G_{n} T_{np}^{dir} + i \int  V_{np}^{dir}  Q_{np} G_{n}
G_{p}T_{np}^{exc},
\label{sec:RBAAM;eq:BSeqnpexc}
\eeqa
where $T_{ij}$ denotes one of the three different
$T$ matrices. The three different bare nucleon-nucleon interactions are
described by one-boson exchange potentials $V_{ij}$. 
In the case of neutrons and protons having different effective
masses, the 
helicity matrix elements cease to be symmetric under the exchange of
these particles leading to an additional sixth independent
helicity matrix element for the $np$ interaction. 
These six independent amplitudes can be reduced to five, if one assumes an average mass 
in the $np$ channel for $V_{np}$.
In that case, $V_{np}^{dir}$  is  related to
$V_{np}^{exc}$ by the Fierz
transformation, which would not be the case otherwise due to the unequal effective masses of neutrons and
protons. Therefore, Eqs.~(\ref{sec:RBAAM;eq:BSeqnpdir})
and~(\ref{sec:RBAAM;eq:BSeqnpexc}) can be reduced to one equation, 
\beqa
T_{np} = V_{np} + i \int  V_{np} Q_{np} G_{n} G_{p} T_{np}.
\label{sec:RBAAM;eq:BSeqnp}
\eeqa
\\ \indent Next, the two-particle propagator $iG_iG_j$ has to be replaced by
the effective propagator. In this work it is the Thompson propagator~\cite{thompson70}. 
The effective Thompson propagator is given by 
\beqa
g_{ij}= i G_{i} G_{j} = \frac{m^*_i}{E^*_i} \frac{m^*_j}{E^*_j}
\frac{1}{\sqrt{s^*}-E^*_i-E^*_j + i \epsilon},
\eeqa
where $\sqrt{s^*}$ is the invariant mass.
\\ \indent Furthermore, the Pauli operator $Q$ has to be replaced by an
angle-averaged Pauli operator $\overline{Q}$. The angle-averaged relativistic Pauli operators for the $nn$ and $pp$
interactions are identical to the one in the symmetric case which is given
in Eq.~(\ref{subsec:SNM;eq:Pauli}). In contrast to the symmetric case,
the Pauli operator for the $np$ interaction has to be evaluated for
Fermi ellipsoids with different sizes. The result for the angle-averaged Pauli operator $\overline{Q}_{np}$  for asymmetric
matter with a neutron excess is  
\beqa
\overline{Q}_{np}=\left\{ \begin{array}{c c c}
\Theta(\gamma u E_{Fn}-\gamma p_{Fn}) & & |\veck|<k_{min} \\
1/2 [\cos(\theta_p)-\cos(\theta_n)] \Theta(\theta_n-\theta_p) & for &
k_{min}<|\veck|<k_{max} \\
1 & & k_{max} < |\veck| \end{array} \right.
\eeqa
with $k_{min}=\gamma | u E_{Fn}- p_{Fn}|$,
$k_{max}=\gamma (u E_{Fn}+ p_{Fn})$, 
\beqa
\theta_p=\left\{ \begin{array}{c c c}
\arccos\left(\frac{\gamma E^*_p(k)-E^*_{Fp}}{\gamma |\veck| |\vecu|}\right) & for &
|\frac{\gamma E^*_p(k) -E^*_{Fp}}{\gamma |\veck| |\vecu|}|  \leq 1 \\
0 & & otherwise \end{array} \right., 
\eeqa
and
\beqa
\theta_n=\left\{ \begin{array}{c c c}
\arccos\left(\frac{E^*_{Fn} - \gamma E^*_n(k)}{\gamma |\veck| |\vecu|}\right) & for &
|\frac{E^*_{Fn} - \gamma E^*_n(k) }{\gamma |\veck| |\vecu|}|  \leq 1 \\
\pi & &  otherwise \end{array} \right. \quad .
\eeqa
However, the central quantities in the model are the neutron self-energy
\beqa
\Sigma_n = -i \int\limits_{F_n} (Tr[G_n T_{nn}] - G_{n}T_{nn}) -i
\int\limits_{F_p} (Tr[G_p T_{np}] - G_{p}T_{np} ),
\label{sec:RBAAM;eq:HFselfeqn1}
\eeqa
and  the proton self-energy
\beqa
\Sigma_p = -i \int\limits_{F_p} (Tr[G_p T_{pp}] - G_{p}T_{pp}) -i
\int\limits_{F_n} (Tr[G_n T_{np}] - G_{n}T_{np} ).
\label{sec:RBAAM;eq:HFselfeqp1}
\eeqa
In Brueckner theory the integrations extend over the Fermi sea of the
neutron $F_n$ and of the
proton $F_p$. Below, the expressions in
Eqs.~(\ref{sec:RBAAM;eq:HFselfeqn1})-(\ref{sec:RBAAM;eq:HFselfeqp1})
will be specified for the different covariant representations.
%
\section{Covariant representation and the Self-Energy Components}
\label{sec:CR&SE}
In this section we will consider  four different
representations of the $T$ matrix: pseudoscalar, conventional
pseudovector, complete pseudovector, and subtracted $T$ matrix
representation.
%
\subsection{Pseudoscalar ($ps$) Representation}
\label{sec:CR&SE:subsec:PS}
The nucleon self-energy components are calculated
in the nuclear matter rest frame using the trace formulas,
Eqs. (\ref{subsec:SM;eq:trace}). Since we determine the $T$ matrix
elements in the two-particle c.m. frame, a representation with  
covariant operators and Lorentz invariant amplitudes 
in Dirac space is the most convenient way to Lorentz-transform the $T$ matrix 
from the two-particle c.m. frame into  the nuclear matter rest frame 
\cite{horowitz87}. On-shell 
a set of five linearly independent covariants is 
sufficient for such a $T$ matrix representation in symmetric nuclear
matter. A linearly independent, however, not unique set of five covariants 
is given by the following Fermi covariants
\beqa
\!\!\!\!\!{\rm S} = 1\otimes 1 ,
{\rm V} =  \gamma^{\mu}\otimes \gamma_{\mu},
{\rm T} = \sigma^{\mu\nu}\otimes\sigma_{\mu\nu}, 
{\rm A} =  \gamma_5 \gamma^{\mu}\otimes \gamma_5 \gamma_{\mu},
{\rm P} = \gamma_5 \otimes \gamma_5 .
\eeqa
To circumvent the problem of unphysical contributions in the nucleon 
self-energy one uses antisymmetrized amplitudes 
$F_i^{\rm I}(\pabs,\theta,x)$ and $F_i^{\rm I}(\pabs,\pi-\theta,x)$. 
Thus, the representation of the $T$ matrix is  
given by \cite{sehn97}
\beq
T^{{\rm I}}(\pabs,\theta,x)=T^{{\rm I},dir}(\pabs,\theta,x)
-T^{{\rm I},exc}(\pabs,\theta,x), \label{ttot}
\eeq
where the ``direct'' and ``exchange'' parts of the $T$ matrix are defined as    
\beqa
T^{{\rm I},dir}(\pabs,\theta,x)&=& {1\over 2} \left[ 
F_{\rm S}^{\rm I}(\pabs,\theta,x){\rm S}+
F_{\rm V}^{\rm I}(\pabs,\theta,x){\rm V}+
F_{\rm T}^{\rm I}(\pabs,\theta,x){\rm T} \right.
\nonumber \\
&+&\left. F_{\rm A}^{\rm I}(\pabs,\theta,x){\rm A}+
F_{\rm P}^{\rm I}(\pabs,\theta,x){\rm P} \right], 
\label{tdir} 
\eeqa
and
\beqa
&&T^{{\rm I},exc}(\pabs,\theta,x)= (-1)^{\rm I+1}{1\over 2}\left[ 
F_{\rm S}^{\rm I}(\pabs,\pi-\theta,x)\tilde{{\rm S}}+
F_{\rm V}^{\rm I}(\pabs,\pi-\theta,x)\tilde{{\rm V}}
\right. \nonumber \\
&&\left. \,\,\,\,\,\,\,\,
+F_{\rm T}^{\rm I}(\pabs,\pi-\theta,x)\tilde{{\rm T}} 
+F_{\rm A}^{\rm I}(\pabs,\pi-\theta,x)\tilde{{\rm A}}
+F_{\rm P}^{\rm I}(\pabs,\pi-\theta,x)\tilde{{\rm P}} \right],
\label{tex}
\eeqa
where the interchanged invariants are defined as~\cite{tjon85a}  $\tilde{{\rm
    S}}=\tilde{{\rm S}} {\rm S}$, $\tilde{{\rm
    V}}=\tilde{{\rm S}} {\rm V}$, $\tilde{{\rm
    T}}=\tilde{{\rm S}} {\rm T}$, $\tilde{{\rm
    A}}=\tilde{{\rm S}} {\rm A}$, and $\tilde{{\rm
    P}}=\tilde{{\rm S}} {\rm P}$ with operator $\tilde{{\rm
    S}}$ exchanging particles 1 and 2, i.e. $\tilde{{\rm
    S}} u(1)_{\sigma} u(2)_{\tau} = u(1)_{\tau} u(2)_{\sigma}$. 
Here $\vecp$ and $\theta$ denote the relative three-momentum and the
scattering angle between the scattered nucleons in the c.m. frame, 
respectively. In addition, the five Lorentz invariant 
amplitudes $F_i^{\rm I}(\pabs,\theta,x)$ with $i=\{{\rm S,V,T,A,P}\}$
depend also on $x=\{\kf, m^*_F,|\vecu|\}$. Taking the single nucleon 
momentum $\veck=(0,0,k)$ along the $z$-axis, we have for the scalar
and vector components of the neutron self-energy 
\beqa
\Sigs (\veck) & = & \frac{1}{4} \int_0^{\kfn} \frac{d^3\vecq}{(2 \pi)^3}  \frac{m^*_n}{E^*_{q,n}} 
[4 F^{nn}_{\rm S} - F^{nn}_{\rm \tilde{S}} - 4 F^{nn}_{\rm \tilde{V}} - 12
F^{nn}_{\rm \tilde{T}} + 4 F^{nn}_{\rm \tilde{A}}   - F^{nn}_{\rm \tilde{P}} ] \nn
\\ & & + \frac{1}{4}
\int_0^{\kfp} \frac{d^3\vecq}{(2 \pi)^3}  \frac{m^*_p}{E^*_{q,p}} 
[4 F^{np}_{\rm S} - F^{np}_{\rm \tilde{S}} - 4 F^{np}_{\rm \tilde{V}} - 12
F^{np}_{\rm \tilde{T}} + 4 F^{np}_{\rm \tilde{A}} - F^{np}_{\rm \tilde{P}}],
\label{subsec:PS;eq:s}
\eeqa
\beqa
\Sigo (\veck) & = & \frac{1}{4} \int_0^{\kfn} \frac{d^3\vecq}{(2 \pi)^3}  
[-4 F^{nn}_{\rm V} + F^{nn}_{\rm \tilde{S}} - 2 F^{nn}_{\rm \tilde{V}} - 2 F^{nn}_{\rm \tilde{A}}   - F^{nn}_{\rm \tilde{P}} ] \nn
\\ & & + \frac{1}{4} 
\int_0^{\kfp} \frac{d^3\vecq}{(2 \pi)^3}  
[- 4 F^{np}_{\rm V} + F^{np}_{\rm \tilde{S}} - 2 F^{np}_{\rm
    \tilde{V}}  - 2 F^{np}_{\rm \tilde{A}} - F^{np}_{\rm \tilde{P}}] ,
\label{subsec:PS;eq:o}
\eeqa
and
\beqa
\Sigv (\veck) & = & \frac{1}{4} \int_0^{\kfn} \frac{d^3\vecq}{(2 \pi)^3}  \frac{\vecq \cdot \veck}{|\veck|^2 E^*_{q,p}}
[-4 F^{nn}_{\rm V} + F^{nn}_{\rm \tilde{S}} - 2 F^{nn}_{\rm \tilde{V}}
  - 2 F^{nn}_{\rm \tilde{A}}  - F^{nn}_{\rm \tilde{P}} ] 
\nn \\ & & + \frac{1}{4} 
\int_0^{\kfp} \frac{d^3\vecq}{(2 \pi)^3} \frac{\vecq \cdot \veck}{|\veck|^2 E^*_{q,p}} 
[- 4 F^{np}_{\rm V} + F^{np}_{\rm \tilde{S}} - 2 F^{np}_{\rm
    \tilde{V}}  - 2 F^{np}_{\rm \tilde{A}} - F^{np}_{\rm \tilde{P}}].
\label{subsec:PS;eq:v}
\eeqa
Corresponding expressions as in
Eqs.~(\ref{subsec:PS;eq:s})-(\ref{subsec:PS;eq:v}) can be obtained for
the components of the proton self-energy. 
\\ \indent This representation gives a strong momentum dependence in
the self-energy in
the symmetric case~\cite{gross99}.
This  strong momentum dependence questions, of course, the validity of the 
'reference spectrum approximation' used in the present self-consistency 
scheme. Furthermore, such a strong momentum dependence leads to unphysical
results deep inside the Fermi sea since the effective mass drops to
values which are close to zero. 
The strong momentum dependence of the self-energy \cite{fuchs98} was found to originate mainly from the one-pion exchange 
contribution to the
self-energy. Therefore, some representations which have a better
treatment of the one-pion exchange contribution are given below.  
%
\subsection{Conventional Pseudovector ($pv$) Representation}
%
As discussed in Sec. \ref{sec:CR&SE:subsec:PS}, 
the set of five covariants used to represent 
the on-shell $T$ matrix is not 
uniquely defined as long as one works exclusively in the subspace 
of positive energy states \cite{poschenrieder88}. 
Obviously, various alternative sets of five linearly
independent covariants exist such as conventional $pv$ representation.
In this representation the pseudoscalar covariant ${\rm
 P}=\gamma_5\otimes \gamma_5$ in the $T$ matrix
representation in Eq.~(\ref{ttot}) is replaced by the pseudovector covariant 
\beqa
{\rm PV} = \frac{\gamma_5 \gamma_{\mu} q^{\mu}}{m^*_i+m^*_j} \otimes \frac{\gamma_5 \gamma_{\mu} q^{\mu}}{m^*_i+m^*_j}
\eeqa
with $q=p_1-p_3$ and $q_0=E_1-E_3$.
It  leads to identical on-shell helicity matrix elements as the 
pseudoscalar covariant. 
\\ \indent Using the $pv$ representation of the $T$ matrix as discussed above 
the nucleon self-energy becomes \cite{terhaar87a,sehn97}
\beqa
\Sigs (\veck) & = & \frac{1}{4} \int_0^{\kfn} \frac{d^3\vecq}{(2 \pi)^3}  \frac{m^*_n}{E^*_{q,n}} 
[4 F^{nn}_{\rm S} - F^{nn}_{\rm \tilde{S}} - 4 F^{nn}_{\rm \tilde{V}} - 12
  F^{nn}_{\rm \tilde{T}} + 4 F^{nn}_{\rm \tilde{A}} \nn
\\ & &  + \frac{m_n^{*2} - k^{* \mu} q^*_{\mu}}{2 m_n^{*2}}  F^{nn}_{\rm \widetilde{PV}} ] + \frac{1}{4}
\int_0^{\kfp} \frac{d^3\vecq}{(2 \pi)^3}  \frac{m^*_p}{E^*_{q,p}} 
[4 F^{np}_{\rm S} - F^{np}_{\rm \tilde{S}} - 4 F^{np}_{\rm \tilde{V}}  \nn 
\\ & &  - 12
  F^{np}_{\rm \tilde{T}} + 4 F^{np}_{\rm \tilde{A}} + \frac{m_p^{*2} +
    m_n^{*2} - 2 k^{* \mu} q^*_{\mu}}{(m_n^*+m_p^*)^2} F^{np}_{\rm \widetilde{PV}}],
\eeqa
\beqa
\Sigo (\veck) & = & \frac{1}{4} \int_0^{\kfn} \frac{d^3\vecq}{(2 \pi)^3}  
[-4 F^{nn}_{\rm V} + F^{nn}_{\rm \tilde{S}} - 2 F^{nn}_{\rm \tilde{V}} - 2 F^{nn}_{\rm \tilde{A}}  \nn
\\ & &   - \frac{E^*_{k,n}}{E^*_{q,n}} \frac{m_n^{*2} - k^{* \mu} q^{*}_{\mu}
}{2 m_n^{*2}}  F^{nn}_{\rm \widetilde{PV}} ] + \frac{1}{4} 
\int_0^{\kfp} \frac{d^3\vecq}{(2 \pi)^3}  
[- 4 F^{np}_{\rm V} + F^{np}_{\rm \tilde{S}} - 2 F^{np}_{\rm \tilde{V}}  \nn
\\ & &  -
  2 F^{np}_{\rm \tilde{A}} -  \frac{2 E^*_{k,n} (m_p^{*2}-k^{* \mu}
q^*_{\mu}) - E^*_{q,p} (m^{*2}_p - m^{*2}_n)}{E^*_{q,p} (m^*_n +
    m^*_p)^2}   F^{np}_{\rm \widetilde{PV}}],  
\eeqa
and
\beqa
\Sigv (\veck) & = & \frac{1}{4} \int_0^{\kfn} \frac{d^3\vecq}{(2 \pi)^3}  \frac{\vecq \cdot \veck}{|\veck|^2 E^*_{q,p}}
[-4 F^{nn}_{\rm V} + F^{nn}_{\rm \tilde{S}} - 2 F^{nn}_{\rm \tilde{V}} - 2 F^{nn}_{\rm \tilde{A}} \nn
\\ & &  - \frac{k_z}{q_z} \frac{m^{*2}_n - k^{* \mu} q^*_{\mu}}{2 m^{*2}_n}  F^{nn}_{\rm \widetilde{PV}} ] + \frac{1}{4} 
\int_0^{\kfp} \frac{d^3\vecq}{(2 \pi)^3} \frac{\vecq \cdot \veck}{|\veck|^2 E^*_{q,p}} 
[- 4 F^{np}_{\rm V} + F^{np}_{\rm \tilde{S}}  \nn
\\ & & - 2 F^{np}_{\rm \tilde{V}} - 2 F^{np}_{\rm \tilde{A}} - \frac{2 k^*_z (m^{*2}_p - k^{* \mu}
  q^*_{\mu} ) - q_z (m^{*2}_p - m^{*2}_n)}{q_z (m^*_n+m^*_p)^2}
F^{np}_{\rm \widetilde{PV}}]. 
\eeqa
The momentum dependence is still pronounced, because the pion contribution is 
not yet treated correctly as pseudovector as discussed in
Ref.~\cite{fuchs98}.
%
\subsection{Complete Pseudovector ($pv$) Representation}
%
\label{CPVR}
Since Fierz transformation mixes the covariants in the conventional $pv$ representation the HF
nucleon self-energy is not reproduced when the pseudovector pion exchange 
potential is used for the bare nucleon-nucleon interaction.
Hence, to suppress the undesired pseudoscalar contribution of the pion to the 
nucleon self-energy we have to determine a different $pv$ representation 
of the $T$ matrix. To obtain a 'complete' $pv$ representation the identities
\beqa
\frac{1}{2} ({\rm T} + {\rm \tilde{T}})= {\rm S} + {\rm
  \tilde{S}} + {\rm P}  + {\rm \tilde{P}}, 
\label{identi1}\\
{\rm V}  + {\rm \tilde{V}} = {\rm S} + {\rm \tilde{S}}  - {\rm
  P}  - {\rm \tilde{P}}
\label{identi2}
\eeqa
are needed.
Applying the operator identities (\ref{identi1})-(\ref{identi2}) the ``symmetrized''
$ps$ representation (\ref{ttot}) can be rewritten as
\beqa
\hspace{1cm}
T^{\rm I} (\pabs,\theta,x)&=& g_{\rm S}^{{\rm I}}(\pabs,\theta,x){\rm S}-g_{\rm \tilde{S}}^{{\rm I}}(\pabs,\theta,x) {\rm \tilde{S}}
+g_{\rm A}^{{\rm I}}(\pabs,\theta,x)({\rm A}-{\rm \tilde{A}}) \nonumber \\ 
&+& g_{\rm P}^{{\rm I}}(\pabs,\theta,x){\rm P}
-g_{\rm \tilde{P}}^{{\rm I}}(\pabs,\theta,x) {\rm \tilde{P}},
\label{tmatrep5}
\eeqa
where the new amplitudes $g_i^{\rm I}$ are defined as 
\beqa
\hspace{2cm}
\left( 
\begin{array}{c} 
g_{\rm S}^{{\rm I}} \\ 
g_{\rm \tilde{S}}^{{\rm I}} \\ 
g_{\rm A}^{{\rm I}} \\ 
g_{\rm P}^{{\rm I}} \\ 
g_{\rm \tilde{P}}^{{\rm I}} \\
\end{array} 
\right)
= {1\over 4}
\left( 
\small{ 
\begin{array}{ccccccc} 
 4 & -2 & -8  & 0  & -2 \\
 0 & -6 & -16 & 0  &  2 \\
 0 & -2 &  0  & 0  & -2 \\
 0 &  2 & -8  & 4  &  2 \\
 0 &  6 & -16 & 0  & -2 \\
\end{array}} 
\right)
\left( 
\begin{array}{c} 
F_{\rm S}^{\rm I} \\ 
F_{\rm V}^{\rm I} \\ 
F_{\rm T}^{\rm I} \\ 
F_{\rm P}^{\rm I} \\ 
F_{\rm A}^{\rm I} \\
\end{array} 
\right) .
\label{transform3}
\eeqa
Due to the linear relations between the amplitudes
$F_i^{\rm I}$ and $g_i^{\rm I}$, the two 
$ps$ representations (\ref{ttot}) and
(\ref{tmatrep5}) of the $T$ matrix lead to identical results 
for the nucleon self-energy.
When we replace in (\ref{tmatrep5}) the covariants 
${\rm P}$, $\tilde{\rm P}$ by the pseudovector covariants ${\rm PV}$, 
$\widetilde{\rm PV}$, respectively, we arrive at the 
complete $pv$ representation \cite{tjon85a}
\beqa
\hspace{1cm}
T^{\rm I}(\pabs,\theta,x)&=& g_{\rm S}^{{\rm I}}(\pabs,\theta,x){\rm S}
-g_{{\rm \tilde{S}}}^{\rm I}(\pabs,\theta,x) {\rm \tilde{S}}
+g_{\rm A}^{{\rm I}}(\pabs,\theta,x)({\rm A}-{\rm \tilde{A}})\nonumber \\
&+& g_{\rm PV}^{{\rm I}}(\pabs,\theta,x){\rm PV}
-g_{\rm \widetilde{PV}}^{{\rm I}}(\pabs,\theta,x) {\rm \widetilde{PV}},
\label{tmatrep6}
\eeqa
with $g_{\rm PV}^{\rm I}(\theta)$ and $g_{\widetilde{\rm PV}}^{\rm I}(\theta)$ 
being identical to $g_{\rm P}^{\rm I}(\theta)$ and 
$g_{\tilde{\rm P}}^{\rm I}(\theta)$, respectively. 
As shown in Ref. \cite{fuchs98}, this representation will be able to reproduce 
the Hartree-Fock results for the nucleon self-energy 
if we use the pseudovector pion exchange potential as bare nucleon-nucleon 
interaction. The self-energy components in the complete 
$pv$ representation of the $T$ matrix are given by 
\beqa
\Sigs (\veck) & = & \frac{1}{4} \int_0^{\kfn} \frac{d^3\vecq}{(2 \pi)^3}  \frac{m^*_n}{E^*_{q,n}} 
[4 g^{nn}_{\rm S} - g^{nn}_{\rm \tilde{S}} + 4 g^{nn}_{\rm A} +
  \frac{m_n^{*2} - k^{* \mu} q^*_{\mu}}{2 m_n^{*2}}  g^{nn}_{\rm
    \widetilde{PV}} ] \nn
\\ & &   + \frac{1}{4}
\int_0^{\kfp} \frac{d^3\vecq}{(2 \pi)^3}  \frac{m^*_p}{E^*_{q,p}} 
[4 g^{np}_{\rm S} - g^{np}_{\rm \tilde{S}} + 4 g^{np}_{\rm A} 
+ \frac{m_p^{*2} + m_n^{*2} - 2 k^{* \mu} q^*_{\mu}}{(m_n^*+m_p^*)^2} g^{np}_{\rm \widetilde{PV}}],
\eeqa
\beqa
\Sigo (\veck) & = & \frac{1}{4} \int_0^{\kfn} \frac{d^3\vecq}{(2 \pi)^3}  
[ g^{nn}_{\rm \tilde{S}} - 2 g^{nn}_{\rm A} - \frac{E^*_{k,n}}{E^*_{q,n}} \frac{m_n^{*2} - k^{* \mu} q^{*}_{\mu}
}{2 m_n^{*2}}  g^{nn}_{\rm \widetilde{PV}} ] \nn
\\ & &   + \frac{1}{4} 
\int_0^{\kfp} \frac{d^3\vecq}{(2 \pi)^3}  
[ g^{np}_{\rm \tilde{S}} - 2 g^{np}_{\rm A} -  \frac{2 E^*_{k,n} (m_p^{*2}-k^{* \mu}
q^*_{\mu}) - E^*_{q,p} (m^{*2}_p - m^{*2}_n)}{E^*_{q,p} (m^*_n +
    m^*_p)^2}   g^{np}_{\rm \widetilde{PV}}],
\eeqa
and
\beqa
\Sigv (\veck) & = & \frac{1}{4} \int_0^{\kfn} \frac{d^3\vecq}{(2 \pi)^3}  \frac{\vecq \cdot \veck}{|\veck|^2 E^*_{q,p}}
[g^{nn}_{\rm \tilde{S}} - 2 g^{nn}_{\rm A} - \frac{k_z}{q_z}
  \frac{m^{*2}_n - k^{* \mu} q^*_{\mu}}{2 m^{*2}_n}  g^{nn}_{\rm
    \widetilde{PV}} ] \nn
\\ & &   + \frac{1}{4} 
\int_0^{\kfp} \frac{d^3\vecq}{(2 \pi)^3} \frac{\vecq \cdot \veck}{|\veck|^2 E^*_{q,p}} 
[g^{np}_{\rm \tilde{S}} - 2 g^{np}_{\rm A} \nn
\\ & & - \frac{2 k^*_z (m^{*2}_p - k^{* \mu}
  q^*_{\mu} ) - q_z (m^{*2}_p - m^{*2}_n)}{q_z (m^*_n+m^*_p)^2} g^{np}_{\rm \widetilde{PV}}].
\eeqa
However,  as already pointed out in \cite{fuchs98},  this
representation will not reproduce HF nucleon self-energy if other
meson exchange potentials than pion exchange potentials are used.
%
\subsection{Subtracted $T$ Matrix Representation}
%
The complete 
$pv$ representation treated in Sect.~\ref{CPVR} will fail to reproduce the HF nucleon self-energy if other 
meson exchange potentials are applied than $\pi$- and ($\eta$-) mesons as bare interaction.
Since the influence of the pion is dominantly given by the
single-pion exchange, it should be reasonable to treat the bare interaction of the
$\eta$ and especially of the pion separately to the rest. After
subtraction of the bare interaction of the $\eta$- and $\pi$-meson $V_{\pi,\eta}$,
the remainder is 
the subtracted $T$ matrix,   
\beq
T_{Sub}=T - V_{\pi,\eta}.
\label{mixed1}
\eeq
The $ps$ representation  should be more
appropriate for the subtracted $T$ matrix because then the higher order contributions of the other 
meson exchange potentials are not treated incorrectly as pseudovector.
Thus one chooses the complete $pv$ representation for $V_{\pi,\eta}$ and  the $ps$
representation for the $T_{Sub}$ to get the most favorable representation.
%
\section{Results and Discussion}
\label{sec:R&D}
\begin{figure}
\begin{center}
\includegraphics[width=0.85\textwidth] {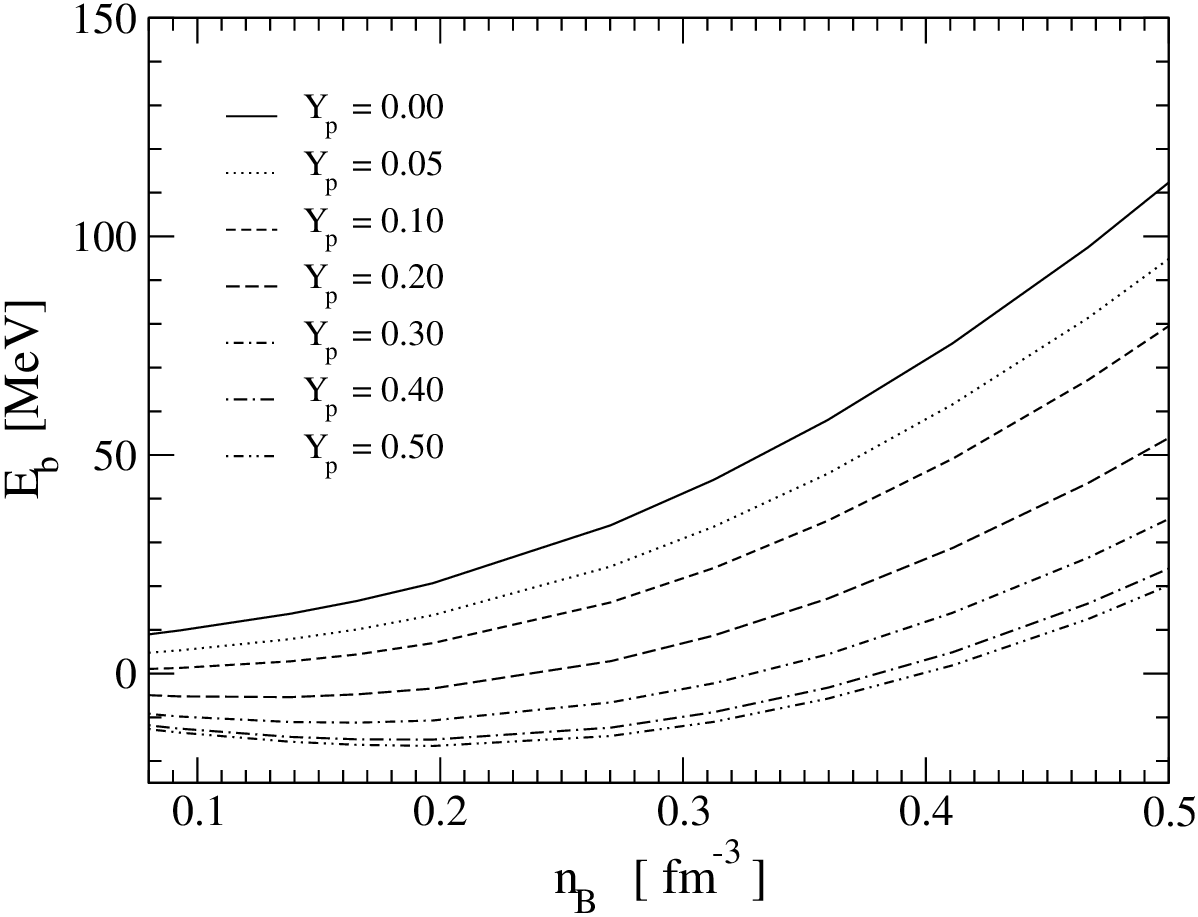}
\includegraphics[width=0.85\textwidth] {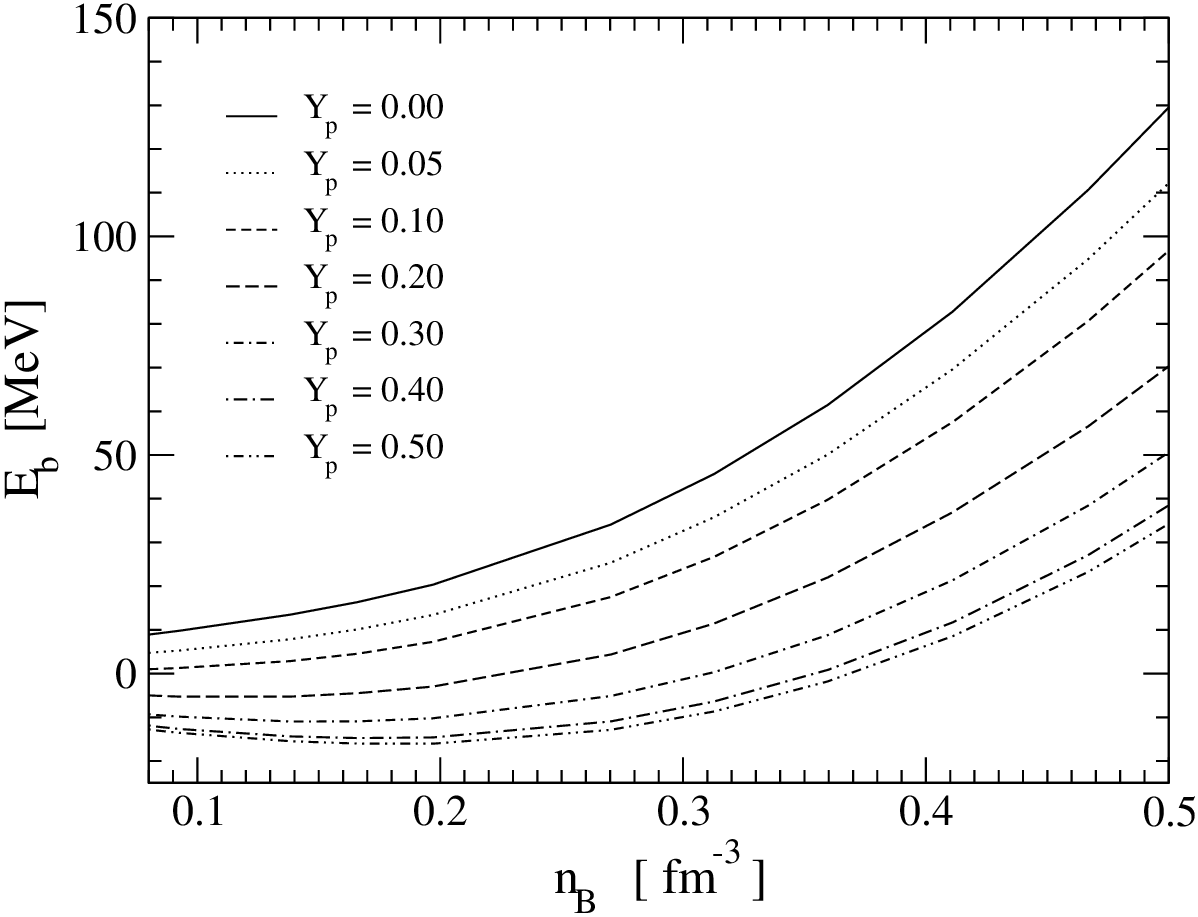}
\caption{Binding energy as a function of the baryon density for proton
  fractions $Y_p$ ranging from 0 to 0.5. The complete $pv$ (top) and
  subtracted $T$ matrix representation (bottom) are used.
\label{fig:BE}}
\end{center}
\end{figure}
In Fig.~\ref{fig:BE} we present the results for the equation of
state for various values of the proton fraction $Y_p=n_p/n_B$ using the Bonn A
potential. The applied representations are the complete $pv$ and the subtracted
$T$ matrix representation. The two extreme cases are symmetric nuclear matter ($Y_p=0.5$) and
neutron matter ($Y_p=0$). The symmetric nuclear matter results agree with those of Ref.~\cite{gross99}.
The binding energy curves for intermediate
values of $Y_p$ lie between these two extreme curves.
In addition to that, the binding energy $E_b$ shows a
nearly quadratic dependence on the asymmetry parameter
$\beta=Y_n-Y_p$ in both representations. Furthermore, the symmetry
energy is defined with help of the binding energy $E_b$ as
\beqa
E_{\rm sym}(n_B)= \frac{1}{2} \left[
  \frac{\partial^2E_b(n_B,\beta)}{\partial \beta^2} \right]_{\beta=0}.
\eeqa
Due to  the nearly quadratic dependence of the binding energy on the
asymmetry parameter, the symmetry energy can be equivalently calculated as
\beqa
E_{\rm sym}(n_B)=E_b(n_B,\beta=1)-E_b(n_B,\beta=0).
\eeqa
\begin{table}[!h]
\begin{tabular}{|c|c|c|}
\hline
representation & $k_F [\textrm{fm}^{-1}]$ & $E_{\rm sym}$ [MeV]  \\
\hline
conventional $pv$ & 1.41 & 30.80\\
complete $pv$ & 1.42 & 36.74\\
subtracted $T$ matrix & 1.39 & 34.36\\
\hline
\end{tabular}
\caption{The symmetry energy at saturation density for various
  representations. The Bonn A potential is used.\label{Esym}}
\end{table}
In Table~\ref{Esym} the calculated symmetry energy is given at saturation density
for each representation.  To compare the results in Table~\ref{Esym}, some values from
literature are given below. The symmetry energy
is found to be 25 MeV using the Groningen potential~\cite{dejong98},
whereas using the Bonn C potential~\cite{machleidt89} it is found to be 28
Mev. In this context it is worth to mention that in Ref.~\cite{dejong98}
the conventional $pv$ representation was used for the decomposition of
the $T$-matrix into Lorentz invariants. The result which the authors
of \cite{dejong98} find for Bonn C is close to the present value
obtained for Bonn A adopting the same projection scheme. However, 
the conventional $pv$ representation suffers from on-shell ambiguities
which lead to large and spurious contributions for the OPE. Thus we
will omit this type of representation in the following. On the other 
hand, in Ref.~\cite{dejong98} a sixth amplitude was introduced which 
appears only in the $np$ channel for the case of 
different neutron and proton masses. For the two limiting cases, 
i.e. symmetric matter and pure neutron 
matter, this amplitude vanishes  identically. Hence the 
results for the symmetry energy are not affected by 
this additional amplitude. As discussed in \cite{dejong98}, the 
definition of this amplitude is, however, not unique. 
Since we use an averaged neutron-proton mass 
for the evaluation of the $V_{np}$ matrix elements as required by 
the Bonn code, we work, in contrast to Ref. \cite{dejong98}, 
standardly with five amplitudes. 
\\ \indent A recent
liquid drop model calculation gives a result of 32.65
MeV~\cite{myers96}. A recent analysis of isovector giant dipole
resonance (GDR) data within relativistic mean-field (RMF) 
theory set a range of $34 \le E_{sym}
\le 36$ MeV~\cite{niksic02}. Hence, in
the complete $pv$ representation the result for the symmetry energy in
Table~\ref{Esym} is probably too high. The calculation, based
on the subtracted $T$ matrix projection scheme, yields a symmetry energy
consistent with  the empirical value. The value of 34 MeV is also in
remarkable agreement with a recent approach to the nuclear many-body
problem based on chiral dynamics in combination with 
QCD sum rules~\cite{finelli04}. 
One has, however, to keep in mind that the saturation density 
obtained with Bonn A is slightly higher than the 
empirical value. Bonn A comes nevertheless closest towards the 
empirical values since Bonn B and C yield smaller densities but 
at the same time significantly too small binding 
energies \cite{brockmann90,gross99}. The comparison performed in 
\cite{finelli04} required therefore a rescaling of the DBHF results in
order to adjust the densities. The symmetry energy does not 
change dramatically, i.e. at $n_B = 0.17~{\rm fm}^{-3}$ the 
values are 33 MeV (subtracted $T$ matrix) and 34 MeV (complete
$pv$).
\begin{figure}
\begin{center}
\includegraphics[width=0.7\textwidth] {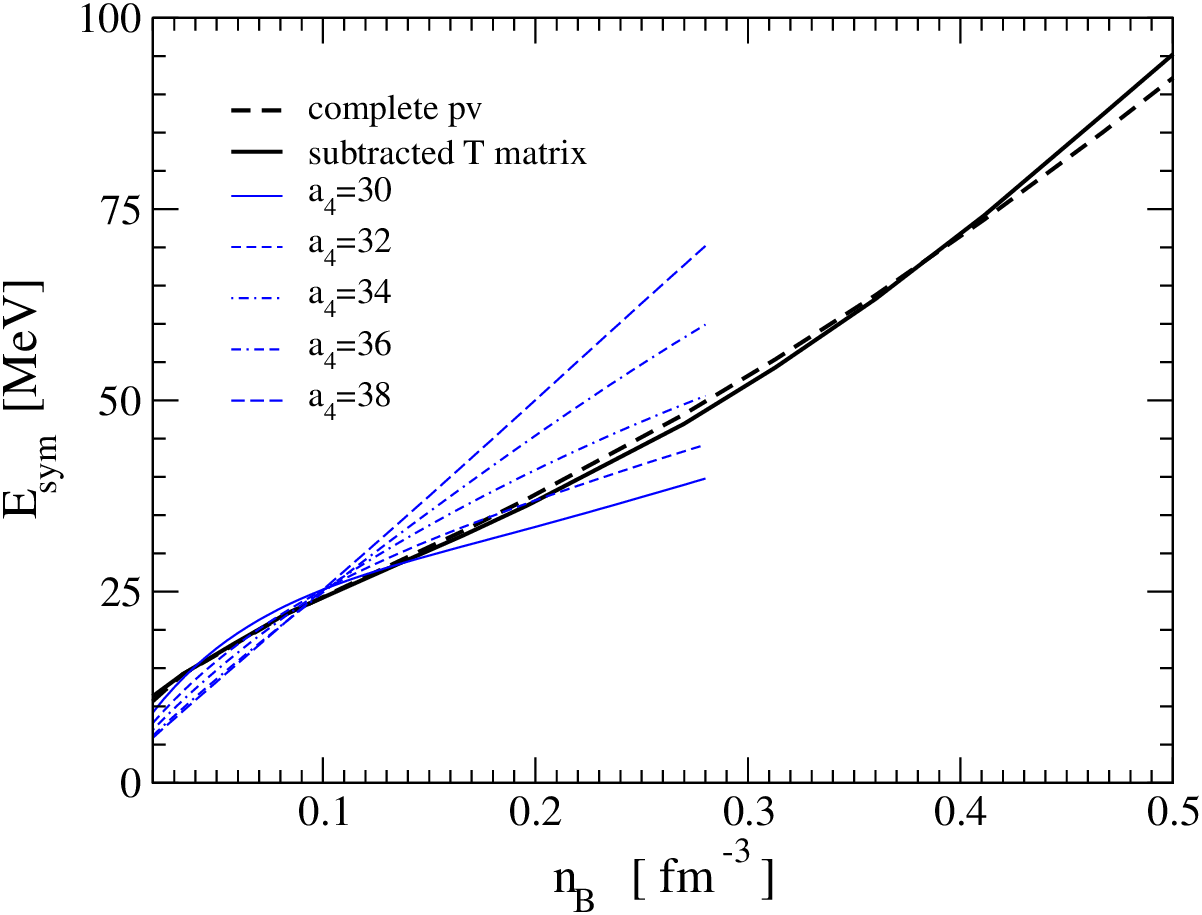}
\caption{Symmetry energy as a function of the baryon density. 
The complete $pv$ and
  subtracted $T$ matrix representation are used.
In addition parameterizations from density dependent 
relativistic mean-field theory \protect\cite{vretenar03} are shown 
where the asymmetry parameter $a_4$ is varied from 30 to 38 MeV.
\label{fig:Esym}}
\end{center}
\end{figure}
\\ \indent The density dependence of the symmetry energy is shown in 
 Fig.~\ref{fig:Esym}. It is similar as, e.g. found in the 
DBHF calculations of \cite{alonso03}.  Although the EOS for 
symmetric matter is relatively soft (K=230 MeV, subtracted $T$ matrix) 
the isospin 
dependence, respectively the symmetry energy, is stiff, in particular at high densities. As generally 
found in relativistic calculations it is significantly stiffer than 
in non-relativistic BHF approaches \cite{bombaci91}. The dependence 
on the representation, i.e. complete $pv$ or subtracted $T$ matrix, 
is weak. For a  quantitative comparison with mean-field 
phenomenology we compare $E_{\rm sym}$ in  
Fig.~\ref{fig:Esym} also to a phenomenological results 
obtained recently within the framework of density dependent 
relativistic mean-field theory \cite{niksic02}. The asymmetry 
parameter $a_4$ is thereby 
varied from 30 to 38 MeV. At moderate densities the dependence of $E_{\rm
 sym}$ is qualitatively similar to the relativistic mean-field
 parameterizations using $a_4 = 32-34$ MeV. However, the density
 dependence of $E_{\rm sym}$ is generally more complex  than in RMF 
theory. In particular at high densities $E_{\rm sym}$ shows a 
non-linear and more pronounced increase. 
For this comparison we have chosen the parameter set 
from \cite{vretenar03} which was obtained at equivalent
 compression moduli (K=230 MeV) although the authors of \cite{vretenar03}
 favor a slightly higher incompressibility for the symmetric case.
\begin{figure}
\begin{center}
\includegraphics[width=0.7\textwidth,] {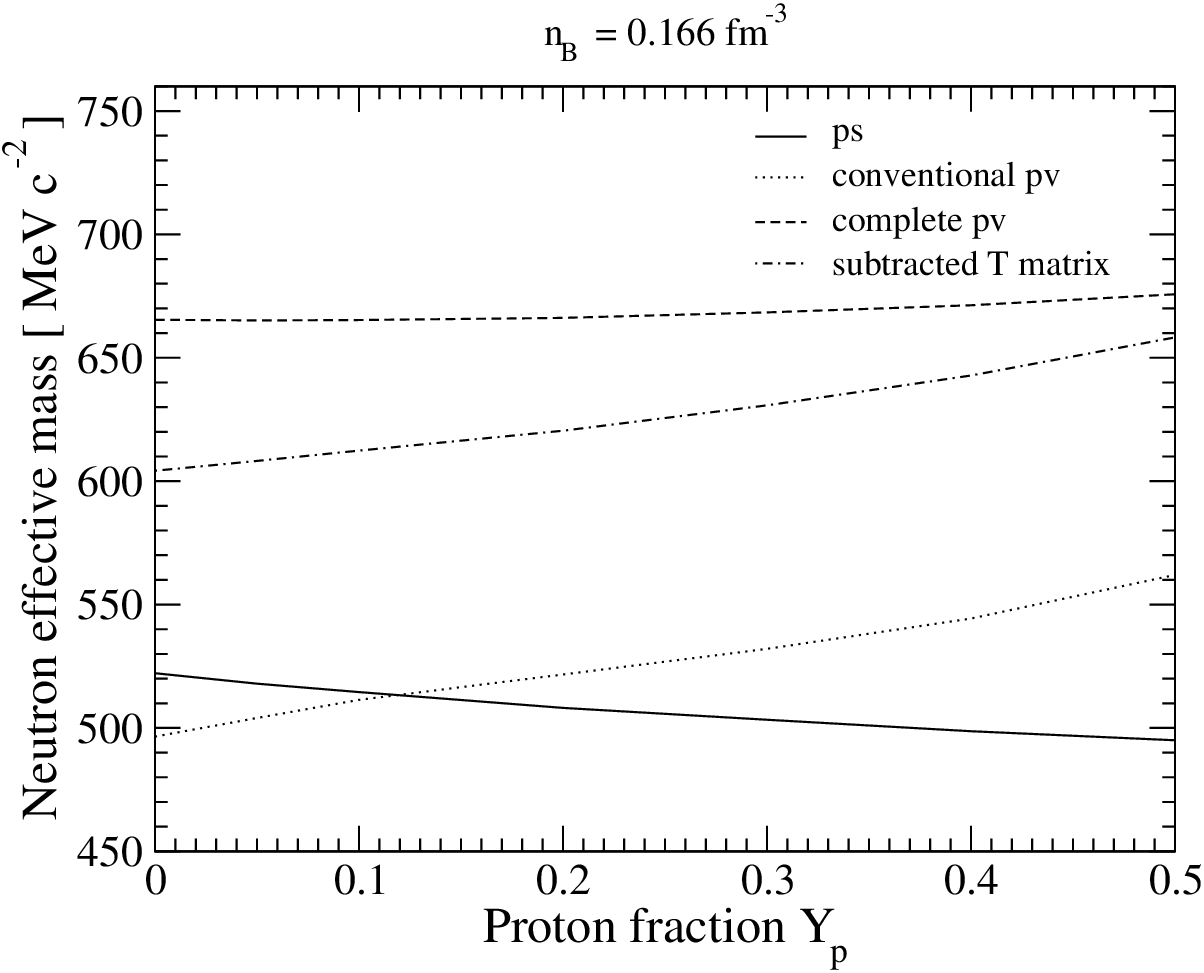}
\caption{Neutron effective mass as a function of the
proton fraction at $n_B = 0.166 \quad \textrm{fm}^{-3}$ for different
representations of the $T$ matrix. 
\label{fig:asymmass}}
\end{center}
\end{figure}
\\ \indent In Fig.~\ref{fig:asymmass} the neutron effective mass is plotted as a
function of the proton fraction $Y_p$ for different
representations at $n_B=0.166 \quad \textrm{fm}^{-3}$. 
As already observed in the symmetric case~\cite{gross99}, the
magnitude of the self-energy, respectively of the effective mass,
depends crucially on the projection scheme which is used for the $T$
matrix. However, observables like the single-particle potential which
are determined by the difference of scalar and vector self-energy
components are much less affected by the different possible projection
schemes since these effects cancel in leading order.
Here we find that the neutron effective mass decreases for the $ps$
representation with increasing proton fraction. This behavior is,
however, an artefact of the $ps$ representation which disappears when
a proper decomposition of the $T$ matrix is used.
Mainly due to a relative large scalar amplitude in the direct part of the  $T$
matrix  in the $nn$ channel compared
to that in the $np$  channel,
the neutron effective mass increases for the conventional
$pv$, the complete $pv$, and the subtracted $T$ matrix
representation with increasing proton fraction. 
\begin{figure}
\begin{center}
\includegraphics[width=0.7\textwidth] {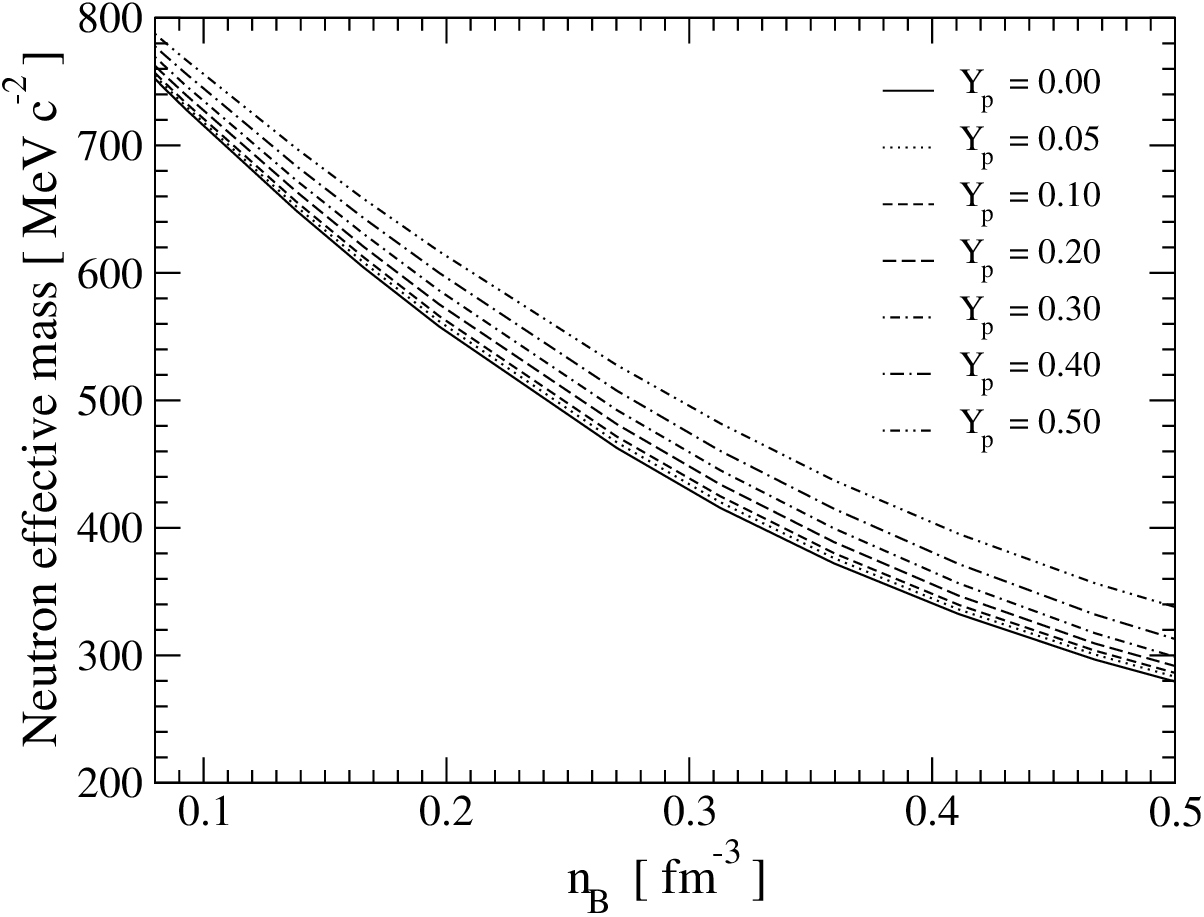}
\caption{Neutron effective mass as a function of the
baryon density for proton
fractions $Y_p$ ranging from 0 to 0.5 using the subtracted
$T$ matrix representation. 
\label{fig:effmass}}
\end{center}
\end{figure}
In Fig.~\ref{fig:effmass} the neutron effective mass is plotted as a
function of the baryon density $n_B$ for various values of the proton 
fraction using the subtracted $T$ matrix representation. Of course, a
strong density dependence can be observed. In addition, the proton fraction
influences the neutron effective mass. 
The upmost curve is the symmetric nuclear matter curve , while the
lowest curve is the neutron matter curve. The symmetric nuclear matter curve
is in agreement with the results of Ref.~\cite{gross99}.
For intermediate
values of the proton fraction $Y_p$ the curve lies between these
two extreme curves.
\begin{figure}
\begin{center}
\includegraphics[width=0.7\textwidth,] {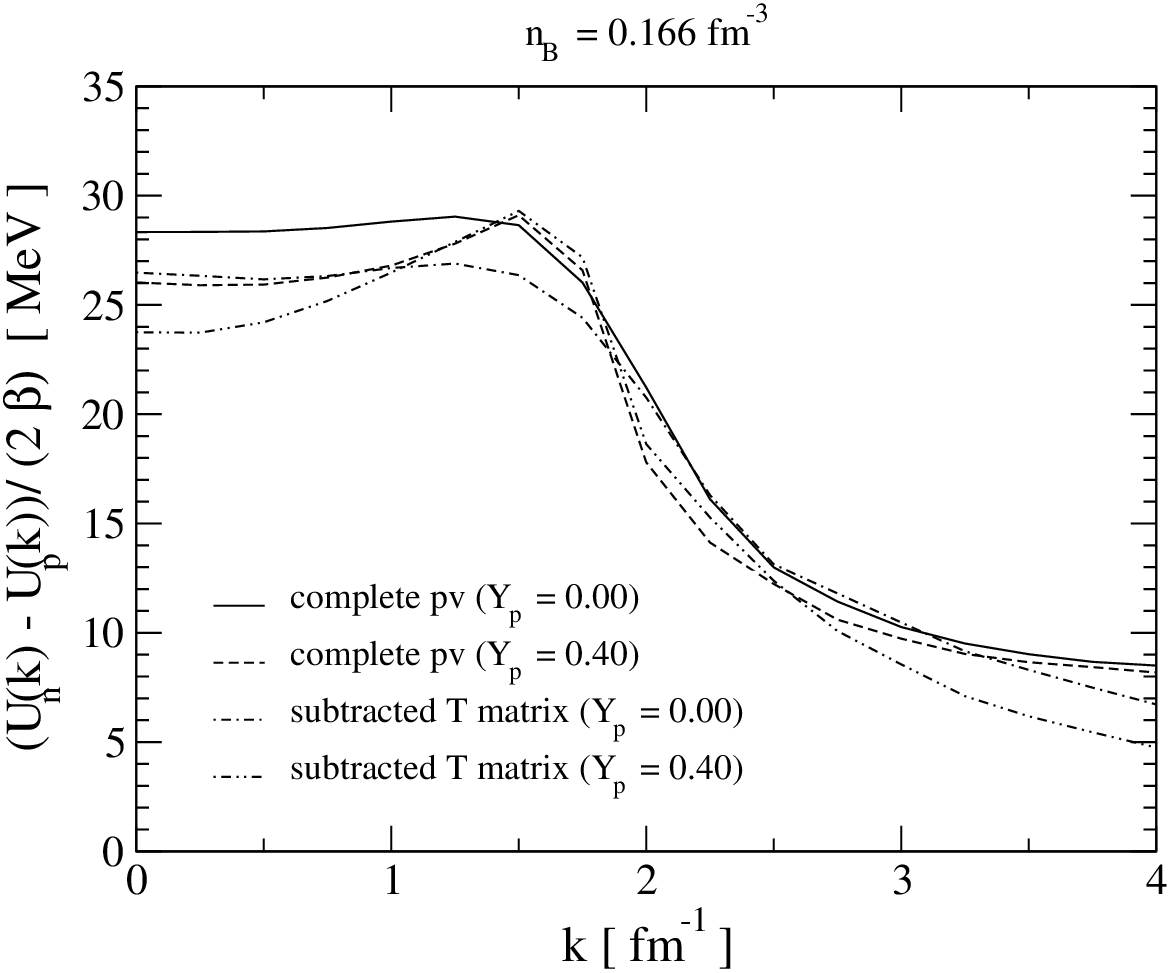}
\caption{Strength of the isovector potential at $n_B = 0.166 \quad \textrm{fm}^{-3}$
as a function of momentum $\veck$~\label{fig:isovectorU}.}
\end{center}
\end{figure}
\\ \indent A quantity which is sensitive on the momentum dependence
of $\Sigma_{s,0,v}$ is the optical potential
which a nucleon feels inside the nuclear medium.
The optical potential is given by
\beqa
U(|\veck|,k^0)=\Sigs(|\veck|)- \frac{1}{M} k^{\mu} \Sigma_{\mu}(|\veck|)
+ \frac{\Sigma_s^2(|\veck|)-\Sigma_{\mu}^2(|\veck|)}{2 M}.
\eeqa
The strength of the isovector nucleon optical potential, i.e., the
Lane potential in~\cite{li04}, can be extracted from $(U_n-U_p)/(2 \beta)$
at saturation density with $\beta=Y_n-Y_p$. From a large number of
nucleon-nucleus scattering experiments at beam energies below about
100 MeV~\cite{patterson76,devito81,deleo81,kwiatkowski78} it can be
concluded that this potential decreases with energy. 
The isovector nucleon optical potentials calculated for the complete $pv$ and subtracted
$T$ matrix representation are shown in Fig.~\ref{fig:isovectorU}.
In these cases the optical potential in neutron rich matter
stays roughly constant  up to a momentum of 1.5 $\textrm{fm}^{-1}$,
corresponding to a kinetic energy of $E_{kin} \sim 45 \quad \textrm{MeV}$, and then
decreases strongly with energy. In mean-field approximation, 
i.e. assuming momentum independent self-energy components, 
the behavior is
different resulting in a continuously increasing optical isovector 
potential~\footnote{We thank M. Di Toro for bringing our attention
towards this fact.}. In the present case the decrease is caused 
by a pronounced explicit momentum dependence of the scalar and vector
isovector self-energy components.
The optical isovector potential
$(U_n-U_p)/(2 \beta)$ at zero momentum is in good agreement with the
empirical value of 22 - 34 MeV~\cite{li04}. Hence, the empirical isovector
potential extracted from proton-nucleus scattering is well reproduced
by the present calculation. The DBHF model predicts thereby
neutron-proton 
effective mass splitting of $m_n^* < m_p^*$. The same behavior
has been found in~\cite{dejong98}. These facts stand in contradiction
with the conclusion drawn in Ref.~\cite{li04} which are based on
nonrelativistic considerations. The same holds if one compares to
nonrelativistic Brueckner calculations for asymmetric nuclear matter
which predict an opposite behavior of the neutron-proton mass
splitting than their relativistic counterparts. In this context one
should, however, be aware that nonrelativistic approaches determine
usually the effective Landau mass, 
\beqa
m^*_{NR} = \left[\frac{1}{M} + \frac{d U}{|\veck| \ d |\veck|
  }\right]^{-1}_{|\veck|=\kf},
\eeqa
 which is in general not
identical with the Dirac mass~(\ref{subsec:SM;eq:dirac}). Only in the limit of a constant, i.e. momentum independent self-energy including
a vanishing spatial component $\Sigv$, both quantities coincide to leading order in the expansion of the relativistic single-particle
energy if - in addition - the nonrelativistic single-particle potential shows a parabolic momentum dependence. However, the relativistic
self-energy is in general momentum dependent and also the nonrelativistic potential is more complex~\cite{baldo00}.   
\\ \indent  The mean-field
effective coupling constants can be obtained from the scalar and
vector self-energies. Such coupling functions parameterize the correlations of the 
T-matrix in a handable way. They can be applied to finite nuclei 
within the framework of relativistic density dependent mean-field 
theory \cite{fuchs95}. The four
channels are: Dirac scalar isoscalar, Dirac vector isoscalar, Dirac scalar
isovector, and Dirac vector isovector channel. The neutron and proton
self-energies, respectively,  are calculated at their Fermi momentum.
Effective coupling constants are then given by  
\beqa
(\frac{g_{\sigma}}{m_{\sigma}})^2 = - \frac{1}{2}
\frac{\Sigsn(p_{Fn})  + \Sigsp(p_{Fp})}{n^s_n + n^s_p}, \\  
(\frac{g_{\omega}}{m_{\omega}})^2 = - \frac{1}{2}
\frac{\Sigon(p_{Fn})  + \Sigop(p_{Fp})}{n^v_n + n^v_p}, \\  
(\frac{g_{\delta}}{m_{\delta}})^2 = - \frac{1}{2}
\frac{\Sigsn(p_{Fn})  - \Sigsp(p_{Fp})}{n^s_n - n^s_p}, \\
(\frac{g_{\rho}}{m_{\rho}})^2 = - \frac{1}{2}
\frac{\Sigon(p_{Fn})  - \Sigop(p_{Fp})}{n^v_n - n^v_p},     
\eeqa
where $n^s_i$ and $n^v_i$ are the scalar and vector densities~\cite{serot86}. 
\begin{figure}
\begin{center}
\includegraphics[width=0.99\textwidth,] {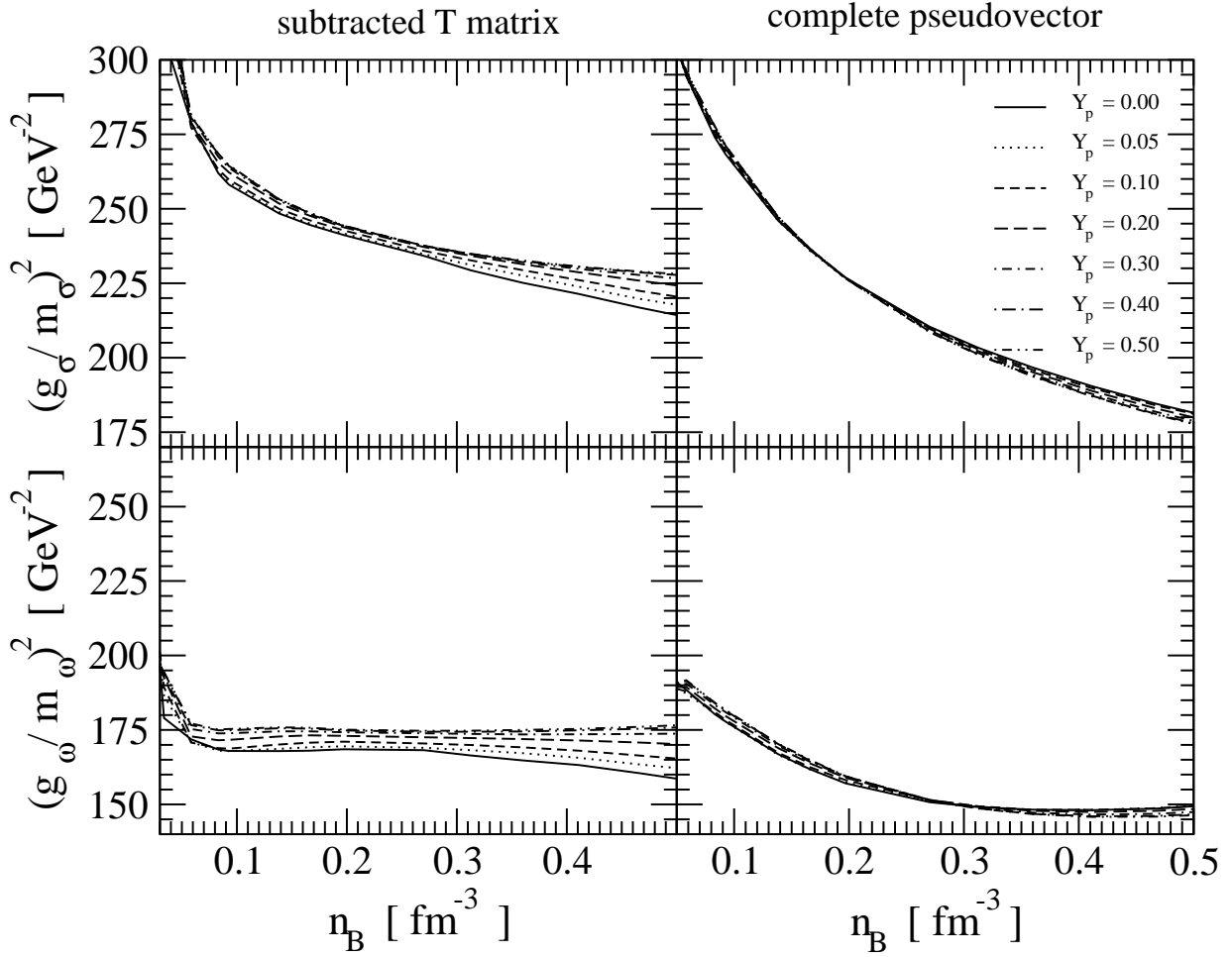}
\caption{The effective coupling constants in the isoscalar scalar $g_\sigma$
  and vector $g_\omega$ as a function of the
baryon density for proton
fractions $Y_p$ ranging from 0 to 0.5 using the complete $pv$ and
  subtracted $T$ matrix representation.  
\label{fig:sigmaomega}}
\end{center}
\end{figure}
\begin{figure}
\begin{center}
\includegraphics[width=0.99\textwidth,] {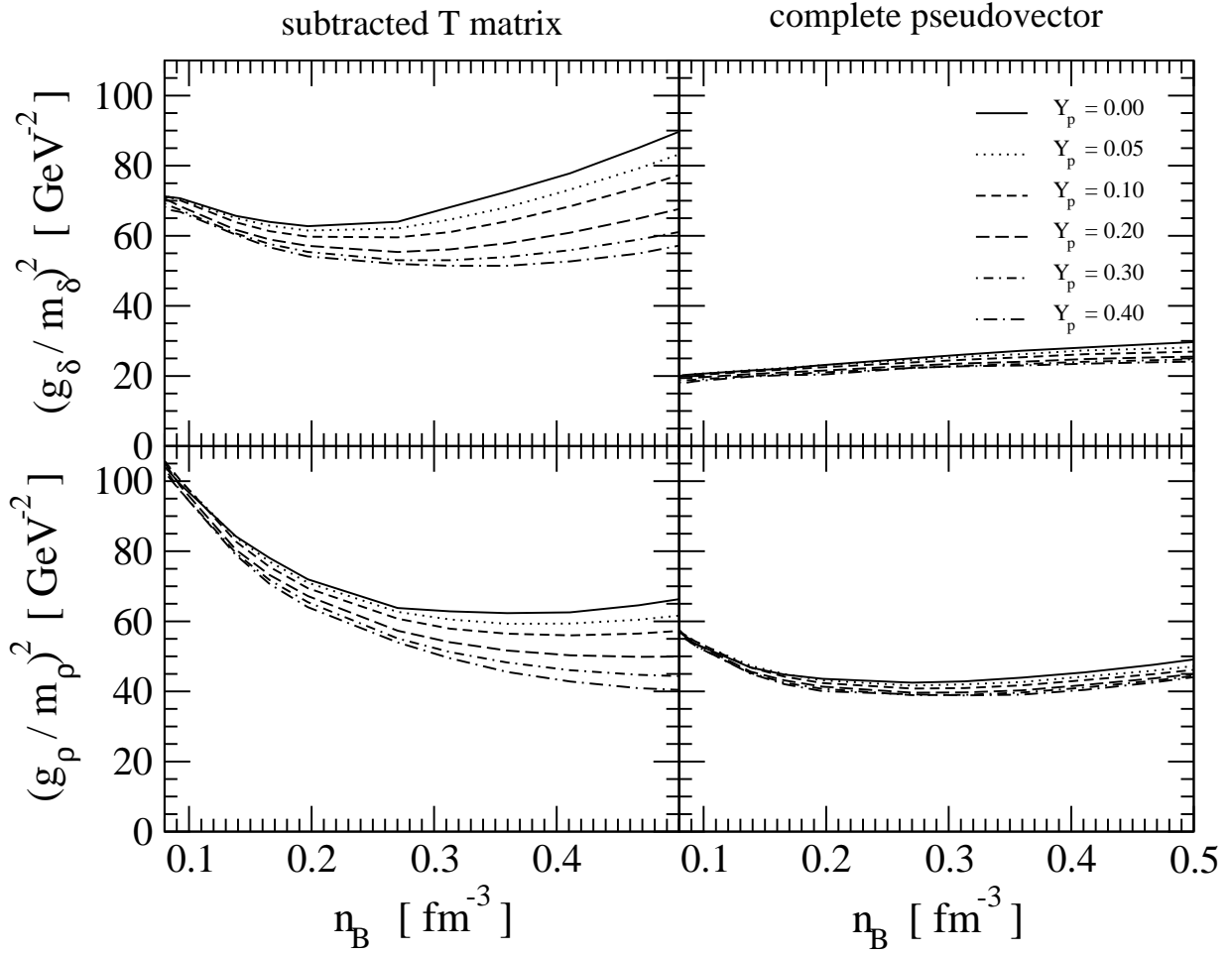}
\caption{The effective coupling constants in the isovector scalar $g_\delta$
  and vector $g_\rho$ as a function of the
baryon density for proton
fractions $Y_p$ ranging from 0 to 0.4 using the complete $pv$ and
  subtracted $T$ matrix representation.  
\label{fig:deltarho}}
\end{center}
\end{figure}
The results for the isoscalar and isovector coupling constants are
plotted in Figs.~\ref{fig:sigmaomega} and~\ref{fig:deltarho},
respectively. The representations used are the complete $pv$
representation and
subtracted $T$ matrix representation. As already mentioned before, the
magnitude of the self-energy depends crucially on the projection
scheme which is used for the $T$ matrix. As a consequence the coupling
constants are also depending on the projection scheme. For the complete $pv$
representation the strength of the isoscalar coupling constants clearly
decreases as the density increases. Applying the subtracted $T$ 
matrix representation - which is the more reliable method - 
the density dependence is less pronounced. At low densities, both the scalar 
$g_\sigma$  and the vector coupling $g_\omega$ show a strong decrease
with densities but then the vector coupling stays more or less
constant. Such a behavior is qualitatively similar to 
relativistic mean-field lagrangians which include non-linear $\sigma$ 
self-energy terms. When cast into the form of density dependent 
couplings these models have a slightly decreasing 
scalar and a constant vector coupling. However, when applied to finite 
nuclei the DBHF isoscalar couplings should be renormalized in 
order to shift the too large saturation density ($0.18~{\rm fm}^{-3}$
with Bonn A) to the phenomenological value. After such a procedure 
the isoscalar DBHF self-energy components are in remarkable agreement 
with the results of \cite{finelli04} based on ChPT + QCD sum rules. 
A renormalization of the DBHF results can be motivated by 
higher order corrections of the hole-line expansion, 
such as e.g. contributions from 3-body forces which are 
known to shift the saturation density towards smaller values
\cite{zuo99}. 
\\ \indent If DBHF could perfectly be mapped on mean-field phenomenology 
the isoscalar coupling functions should 
not depend on the isovector parameter. Since the T-matrix is a 
complex object where isovector and isoscalar contributions mix 
such an idealization can not be expected to hold completely. 
However, the influence of the density on
the coupling constants is much larger than that of the proton
fraction. For the subtracted $T$ matrix representation,
the isoscalar coupling constants show also some dependence  on
the proton fraction. However, here as well as in the work in
Ref.~\cite{dejong98} the dependence of the coupling constants on the proton
fraction is generally weak. For the subtracted $T$ matrix
representation the isoscalar vector channel shows the strongest 
variation as a function of the proton fraction $Y_p$ which lies in
this case between 5 and 10 \%.  
\\ \indent The strength in the isovector channel is small
compared to that in the isoscalar channel, especially for the complete $pv$
representation. Furthermore, 
 a strong dependence on the proton fraction is observed for the
 subtracted $T$ matrix representation. 
The results for the 
isovector channel in case of the subtracted $T$ matrix representation 
lie closest to the results of Ref.~\cite{dejong98} which are based on
the Groningen $NN$ interaction. In particular we observe the same 
increase of the scalar isovector coupling $g_\delta$ at
large densities. The variation of the isovector couplings  as a 
function of the proton fraction $Y_p$ is generally somewhat larger 
than observed in \cite{dejong98}. 
%
\section{Summary and Conclusion}
\label{sec:c}
We present a calculation of asymmetric nuclear matter in a
relativistic Brueckner framework using the Bonn A potential. 
The standard treatment, in which the positive-energy-projected on-shell
$T$ matrix  has to be decomposed into Lorentz invariant
amplitudes, is applied. Furthermore, the $T$ matrix is
represented by a set of five linearly independent Lorentz invariants
with an average mass for the $np$ interaction. 
Since the restriction to positive energy states causes on-shell 
ambiguities concerning the pseudoscalar
or pseudovector nature of the interaction~\cite{huber95,dejong98}, some freedom in the choice of the
representation exists and  the relativistic nuclear self-energy can not be
determined in a unique
way.  This ambiguity is minimized
by separating the leading order, i.e. the single-meson exchange, from
the full $T$ matrix. Therefore,  the contributions
stemming from the single-$\pi$ and-$\eta$ exchange are given in the complete
$pv$ representation. For the other
single-meson exchanges and  the  remaining higher
order correlations, the $ps$ representation is chosen. 
\\ \indent With this representation scheme, the subtracted $T$
representation scheme~\cite{gross99},  we have performed calculations for several
observables in asymmetric nuclear matter and compared them with the
results for other representation schemes, e.g. the complete $pv$
representation scheme. The binding energy shows the expected 
quadratic dependence on the
asymmetry parameter in both representations. The symmetry energy
is found to be 34 MeV at saturation density, whereas in the complete
$pv$ representation the value of 37 MeV is found, which is too high.  
With increasing proton fraction the neutron effective mass increases. 
In addition, the density dependence 
is stronger than the asymmetry dependence on the neutron effective
mass.  Furthermore, a mass
splitting of $m_n^* < m_p^*$ is found in both cases.
\\ \indent In summary, Bonn A predicts a nuclear equation of state which 
is soft for symmetric matter (K=230 MeV) but has a stiff isospin 
dependence ($E_{\rm sym} = 34$ MeV). Both facts are in agreement 
with constraints obtained from heavy ion reactions where kaon 
production can be used to constrain the symmetric part \cite{fuchs01} and pion production \cite{gaitanos04}
and isospin diffusion \cite{chen04} to constrain the symmetry energy .  
\\ \indent 
The results are also analyzed in terms of mean-field Dirac
scalar-vector isoscalar-isovector quantities.
In general the strength in the isovector channel is small
compared to that in the isoscalar channel. However, the found
strength in the isovector channel is still significant for the subtracted $T$ matrix. Furthermore, a strong
dependence on the proton fraction is observed in that case.
\\ \indent Therefore, the results above have consequences for nuclei
with large neutron excess and for neutron stars. 
%
%

\end{document}